\documentclass[useAMS,usenatbib]{mn2e}
\usepackage[dvips]{graphicx}
\usepackage{natbib}
\usepackage{hyperref}
\usepackage{deluxetable}
\usepackage{amssymb}
\title[Population Synthesis of Isolated Neutron Stars with Magneto--rotational Evolution]
{Population Synthesis of Isolated Neutron Stars with Magneto--rotational Evolution}

\author[M.~Gull\'on et al.]{Miguel Gull\'on$^1$, 
  Juan A. Miralles$^1$, Daniele Vigan\`o$^2$, Jos\'e A.~Pons$^1$ \\ 
$^1$ Departament de F\'isica
  Aplicada, Universitat d'Alacant, Ap. Correus 99, 03080 Alacant,
  Spain\\ 
$^2$ Institute of Space Sciences (CSIC--IEEC), Campus UAB,
  Faculty of Science, Torre C5-parell, E-08193 Barcelona, Spain}

\begin{document}

\date{Accepted 00 XXXX 00. Received 00 XXXX 00; in original form 00 XXXX 00}

\pagerange{\pageref{firstpage}--\pageref{lastpage}} \pubyear{}

\maketitle

\label{firstpage}

\newpage

\begin{abstract}

We revisit the population synthesis of isolated radio-pulsars incorporating 
recent advances on the evolution of the magnetic field and the angle between the magnetic and rotational axes
from new simulations of the magneto-thermal evolution and magnetosphere models, respectively.
An interesting novelty in our approach is that we do not assume the existence of a death line.
We discuss regions in parameter space that are more consistent with the observational 
data. In particular, we find that any broad distribution of birth spin periods with $P_0\lesssim 0.5$ s can fit the data, and that
if the alignment angle is allowed to vary consistently with the torque model, realistic magnetospheric models are favoured 
compared to models with classical magneto-dipolar radiation losses.
Assuming that the initial magnetic field is given by a lognormal distribution, our optimal model has mean strength 
$\langle\log B_0{\rm [G]}\rangle \approx 13.0-13.2$ with width $\sigma (\log B_0) = 0.6-0.7$. However, there are strong
correlations between parameters. This degeneracy in the parameter space can be broken by an independent estimate 
of the pulsar birth rate or by future studies correlating this information with the population in other observational bands (X-rays and $\gamma$-rays). 
\end{abstract}


\section{Introduction}

Almost fifty years after the discovery of the first radio--pulsar, the increasing wealth of data provided by neutron star
(NS) observations in the whole electromagnetic spectrum has led to a variety of classes that should be explained by a unified theoretical model. Population synthesis is a useful method to discriminate among different theoretical models by 
means of a statistical analysis. The method requires a large number of observed objects in order to reduce the statistical errors in the determination of the parameters of the underlying model. Nowadays, the number of observed radio-pulsars has increased enough to allow a number of studies \citep{Gon2004,Faucher,PSrev,Bates2014,Szary2014}. This approach can be also extended to the population of NSs observed in other bands (e.g., X-rays, \citealt{PSB_Popov}, $\gamma$-rays, \citealt{Pierbattista}), as statistics increase with the improved sensitivity of new instruments.

The most advanced simulations of the magneto-thermal evolution \citep{Vigano2013} and improved magnetospheric models \citep{Philippov} are now able to predict more precisely the long-term behaviour of magnetic field strength, angular momentum loss rate and the angle between the magnetic dipole moment and the rotational axis. This latter issue is of particular interest, in the light of 
the observational evidence of the variability of this angle \citep{WJ2008,Young2010,Lyne2013}. In addition, the most recent evolutionary models 
of the long term evolution of NSs point to an important results: there is substantial magnetic field decay, consistent with the results in the seminal work of \cite{Gon2004}
and with the observed distribution of periods of isolated X-ray pulsars \citep{Pons2013}. Thus, a revision or extension
of previous studies of population synthesis (PS) of isolated NSs is now timely.   
Our main aim is to reproduce the observed populations in different bands within a unified evolution model, and constrain the allowed regions in the parameter space of the initial distribution of period and magnetic fields at birth.
In this first paper, we focus on radio-pulsars, using the Parkes Multibeam (PMB, \citealt{pksmb}) and Swinburne Multibeam (SMB, \citealt{swmb}) radio surveys, including the pulsars found in their later extensions \citep{Burgay2006, Jacoby2009}. In following works, we will combine the population analysis and synthesis of the radio pulsars with the pulsar distributions detected in other observational bands. The paper has the following structure. In \S~\ref{sec:obs}, we describe the observational data samples employed in the analysis. In \S~\ref{sec:ps} we introduce the PS code and discuss the magneto--rotational evolutionary models. \S~\ref{sec:disc} is devoted to discuss the results, while \S~\ref{sec:conc} is left for the summary.

\section{Observational samples}\label{sec:obs}

More than two thousand radio-pulsars have been registered in the ATNF database\footnote{http://www.atnf.csiro.au/research/pulsar/psrcat/} \citep{ATNF} and about $30$ radio surveys have been performed during several decades. Different surveys cover different regions of the Galaxy, at different frequencies, and/or with different sensitivities, which results in observational biases that introduce much dispersion in the complete pulsars catalogue.
Therefore, the comparison of theoretical models with the distributions of the complete sample of all detected pulsars is not trivial. For this reason, in this paper we consider only objects in the PMB \citep{pksmb} and SMB \citep{swmb} surveys.
They have the same central frequency ($\nu = 1400$ MHz) and most of the detection parameters 
(band width, gain, receiver temperature).
On the other hand, the covered regions of the sky are complementary:
PMB covers the Galactic disk ($|b| \lesssim 5^\circ$), while SMB covers intermediate latitudes ($5^\circ \lesssim b \lesssim 15^\circ$).
The longitude range is the same, $l \in [260^\circ, 50^\circ]$.
The Parkes High-Latitude pulsar survey (PHL, \citealt{Burgay2006}) extends the coverage, using the same observing system, 
to longitudes in the range $l \in [220^\circ, 260^\circ]$ and up to latitudes $|b| \lesssim 60^\circ$.
PHL detected 42 pulsars of which 18 were new discoveries.
The SMB survey has also been extended \citep{Jacoby2009} with the discovery of 26 new pulsars including seven binary and/or millisecond pulsars.
For our purpose in this work, 
we exclude from the analysis the radio-pulsars with $\dot{P} < 0$, and those with period $P<30$ ms, likely to be recycled pulsars.
This results in a population of $1206$ isolated radio-pulsars
($N_{\rm pmb}=1008$ in PMB, $N_{\rm smb}=197$ in SMB, $N_{\rm phl} = 31$ in PHL, with 23 and 7 common detections in PMB-SMB and PMB-PHL respectively).

An important parameter is the threshold flux, $S_{\rm min}$, above which a pulsar can be detected. It strongly depends on the position of the star, because this determines the sky temperature in the pulsar direction and the electron column density between the pulsar and the Earth. It also depends on the survey parameters and, more weakly, on the spin period of the pulsar and its pulse width \citep{Dewey}. 

\begin{figure}
\begin{center}
\includegraphics[width=8cm]{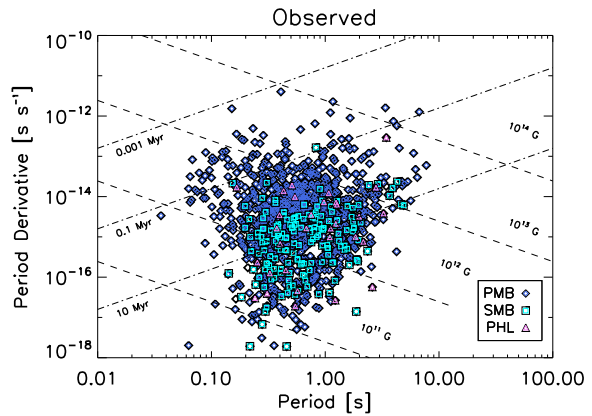}
\end{center}                                                                        
\caption{$P$-$\dot{P}$ diagram for the pulsars detected in the PMB, SMB and PHL surveys. 
Dashes represent isocontours of the inferred magnetic strength at the pole, using the common formula $B = 6.4 \times 10^{19} (P{\rm[s]} \dot{P})^{1/2}$ G. 
Dash-dotted lines correspond to isocontours of constant characteristic age, $\tau = P/2 \dot{P}$. }
\label{fig:ppdotcat}
\end{figure}

Fig. \ref{fig:ppdotcat} shows the observed pulsar distribution in the $P$-$\dot{P}$ diagram for the PMB, SMB and PHL samples.
In both samples, values of $P$ and $\dot{P}$ are very disperse and span about two and five orders of magnitude, respectively. In Table~\ref{tab:surveys} we quantify the statistical differences between PMB and SMB samples, showing the $p$-values obtained with a 1-D Kolmogorov-Smirnov significance test (KS test) for the primary magnitudes $P$ and $\log(\dot{P})$ and for the distances to the Earth given by the ATNF catalog. 
We note that the period distributions are compatible, but the mean values of $\dot{P}$ differ by one order of magnitude, with PMB pulsars being typically younger (i.e., larger $\dot{P}$) than those in the SMB sample. Such difference is consistent with  
most pulsars being born close to the Galactic plane (covered by PMB). As the population gets older, due to the kick velocity at birth, they can move towards higher latitudes (covered by SMB) with timescales comparable to their characteristic ages.

\begin{table}
\begin{tabular}{|c | c | c | c |}
\hline 
\hline   
Magnitude 	& PMB & SMB & $p$-value \\
\hline
$P$	(s)	&	$0.82 \pm 0.84$	&	$0.83 \pm 0.78$	&	$0.19$\\
$\log(\dot{P})$	&	$-14 \pm 1$	&	$-15.1 \pm 0.8$	&	$2.2 \times 10^{-17}$	\\
$d$ (kpc)	&	$7	\pm 6$	&	$5 \pm 3$ 	&	$1.7 \times 10^{-13}$	\\
\hline
\hline
\end{tabular}
\caption{Mean values (with standard deviation) of the distributions of $P$, $\log(\dot{P})$ and distance to the Earth, $d$, in the PMB and SMB surveys. The associated significance $p$-value from the KS test is also shown.}
\label{tab:surveys}
\end{table}

\section{Population synthesis}\label{sec:ps}

We performed Monte Carlo simulations to generate synthetic samples of ``observed pulsars", considering their physical evolution and the observational selection effects. To this purpose, we used an upgraded version of the code used in \S~5 of \cite{PSB_Popov}, partly following the methodology employed by previous authors \citep{Faucher}.
In the following we review the most important parameters and describe the main novelties incorporated in the simulations: the time-dependent evolution of magnetic field and the angle between the magnetic and rotational axis (hereafter denoted by $\chi$).

\subsection{Age, birth and kinematics}

We generate samples with random ages uniformly distributed in the interval $t \in [0,t_{\rm max}]$, where we take $t_{\rm max}=560$ Myr.
 Results are insensitive to the exact value of $t_{\rm max}$, as long as it is large enough to safely assume that NSs older than $t_{\rm max}$ are not detected.
This is equivalent to assume a constant birth rate:
\begin{equation}\label{eq:birthrate}
n_{\rm br} = \frac{N_{\rm star}}{t_{\rm max}}~,
\end{equation}
where $N_{\rm star}$ is the total number of generated stars. The different physical and observational selection effects allow the detection of only $N_{\rm det}$ pulsars, a small fraction of $N_{\rm star}$. We stop each simulation when the numbers of synthetic pulsars detectable by PMB matches the observed number (1008). 

The NSs are thought to be born mainly within the Galactic spiral arms, as these regions are rich in OB stars.
For this purpose we use the model of the Galactic structure by \cite{Milky_way}.
The initial positions are randomly obtained according to the radial probability distribution of \cite{radial},
and the following dependence on the height from the Galactic equatorial plane $z_0$:
\begin{equation}
p(z_0) = \frac{1}{\langle z_{0} 
\rangle} \exp{\left(-\frac{|z_0|}{\langle z_{0} \rangle}\right)}~.
\end{equation}
NSs are assumed to be born with an initial velocity $v_0$, which distribution is also assumed to be exponential:
\begin{equation}
p(v_0) = \frac{1}{\langle v_{0} \rangle} \exp{\left(-\frac{v_0}{\langle v_{0} \rangle}\right)}~.
\end{equation}
The direction of the velocity vector is randomly chosen with equal probability in all directions.

The position of each NS is then evolved until its present age $t$, by solving the Newtonian equations of motion under the influence of the smooth Galactic gravitational potential \citep{galpot1,galpot2}. We do not consider the rotation of the Galaxy, which would smear out the effect of the spiral arm structure for the oldest pulsar population (with ages comparable with $\sim 10^8$ yr, the Galactic rotation period. This, and other corrections due to local structures (e.g. the local arm), will be incorporated in future work. The mean Galactic height $\langle z_{0} \rangle$ and the mean velocity $\langle v_{0} \rangle$ determine the dispersion of the initial and evolved distributions in height.

\begin{figure}
\begin{center}
\includegraphics[width=8cm]{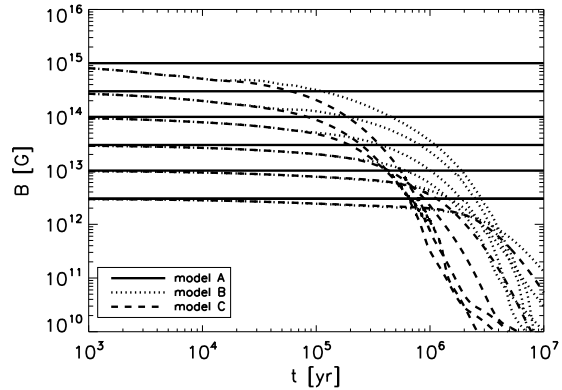}
\end{center}
\caption{Evolution of the strength of the dipolar component of the magnetic field at the pole, $B(t)$, according 
to the magneto-thermal simulations. Solid lines correspond to the constant magnetic field case (models A),
dotted lines show the curves for the intermediate case (models B), and 
dashed lines correspond to the model with fast field decay  (models C).}
\label{fig:curves}
\end{figure}

\subsection{Evolution of spin period, magnetic field, and obliquity.}

As NSs age, they spin down because their rotational energy is lost due to magnetospheric torques \citep{Beskin, Philippov}, which depend on the external dipolar component of the magnetic field. We assume Gaussian distributions for both the initial spin period $P_0$ and the logarithm of the initial magnetic field, $B_0$, that is,
\begin{equation}
\begin{array}{ll}
p(P_{0}) = \frac{1}{\sqrt{2 \pi} \sigma_P0} \exp{\left(-\frac{(P_{0} - \mu_{P_0})^2}{2 \sigma_{P_0}^2}\right)}~, \\
p(\log{B_{0}}) = \frac{1}{\sqrt{2 \pi} \sigma_{B_0}} \exp{\left(-\frac{(\log{B_0} - \mu_{B_0})^2}{2 \sigma_{B_0}^2}\right)}~. \\
\end{array}
\end{equation}
Negative periods are rejected. The initial angle, $\chi_0$, is chosen in the range $[0,\pi/2]$ according to the probability distribution
\begin{equation}\label{eq:initial_angle}
p(\chi_0) = \sin\chi_0~.
\end{equation}
We then solve the system of equations describing the coupled evolution of $\chi$ and $P$ \citep{Philippov}:

\begin{eqnarray}
&& \dot{\chi} = - \kappa_2 \beta \frac{B^2}{P^2} \sin{\chi} \cos{\chi}~, \label{eq:alignment}\\
&& \dot{P} = \beta \frac{B^2}{P} (\kappa_0 + \kappa_1 \sin^2{\chi})~,\label{eq:spindown}
\end{eqnarray}
where $\beta = \frac{\pi^2 R^6}{I c^3}$, $I$ is the moment of inertia, $R$ the NS radius, $B$ is the magnetic field strength at the magnetic pole and $c$ the speed of light. In our model, NS has a mass of $1.4$ M$_{\odot}$, a radius of $11.6$ km and moment of inertia $I \sim 1.5 \times 10^{45}$~g~cm$^2$, which gives $\beta \sim 6 \times 10^{-40}$ G$^{-2}$. Note that $B$ is, in general, a function of time, $B(t)$, discussed below.

The values for the coefficients $\kappa_1$, $\kappa_2$ determine the magnetospheric torque. We use the sets of values 
obtained from the most recent 3D simulations for vacuum, force-free and resistive magnetospheres \citep{Philippov}. The alignment of the rotation and magnetic axis in pulsars with vacuum magnetospheres is much faster (exponential, with characteristic time $\tau_0 = \frac{P_0}{2 \beta B_0}$) than for plasma-filled magnetospheres (a power-law). We will consider the two cases with constant ($\kappa_2=0$) or evolving ($\kappa_2>0$) angle.

The value $B(t)$ entering in the previous equations reflects the evolution of the magnetic field inside the star. The latter is expected to decay as the initial electric currents supporting the field are dissipated due to the non-negligible resistivity of the NS crust \citep{Aguilera,Pons2009,Vigano2013}. We extract the numerical values $B(t)$ from simulations of the magneto-thermal evolution of NSs \citep{Vigano2013}. In particular, we consider the models with currents circulating in the crust, since they are required to explain the large X-ray luminosities of magnetars. In these models, one of the most relevant microphysical inputs is the charge impurity content of the crust, which determines the magnetic diffusivity and, therefore, the field decay timescale. A relatively large impurity parameter in the inner crust, $Q_{\rm imp}^{\rm ic}$, has been proposed to be the main reason for the observed clustering of periods at $P < 12~$s of isolated X--ray pulsars \citep{Pons2013}. In this work we will study three representative cases: no field decay (model A), moderate field decay on typical timescales of 1 Myr (model B), resulting from simulations with  $Q_{\rm imp}^{\rm ic} = 25$, and the more extreme case with $Q_{\rm imp}^{\rm ic} = 100$, with very fast field decay (model C). In Fig.~\ref{fig:curves} we plot the corresponding functions $B(t)$. In Table~\ref{tab:models} we summarize the five models considered in this work, accounting for the different choices of magnetic field decay, alignment and magnetospheric torques. For practical purposes, we have tabulated the functions $B(t)$ obtained from the simulations and use a logarithmic interpolation procedure to obtain the values at any time during the evolution.

\begin{table}
\begin{center}
\begin{tabular}{c l l c c c}
\hline
\hline
Model & Magnetic field & Magnetosphere & $\kappa_0$ & $\kappa_1$ & $\kappa_2$  \\
\hline
A0	& Constant				& Plasma	& $1$ & $1$ & $0$	\\
B0	& Decay $Q_{\rm imp}^{\rm ic}=25$ 	& Plasma 	& $1$ & $1$ & $0$  \\
C0	& Decay $Q_{\rm imp}^{\rm ic}=100$ 	& Plasma 	& $1$ & $1$ & $0$  \\  
B1	& Decay $Q_{\rm imp}^{\rm ic}=25$ 	& Plasma	& $1$ & $1$ & $1$  \\
B2	& Decay $Q_{\rm imp}^{\rm ic}=25$ 	& Vacuum 	& $0$ & $\frac{2}{3}$ & $\frac{2}{3}$ \\
\hline
\hline
\end{tabular}
\end{center}
\caption{Magneto-rotational models considered in this work. The models with magnetic field decay consider different magneto-thermal evolution models, with the indicated values of values of the impurity parameter in the innermost part of the crust, $Q_{\rm imp}^{\rm ic}$ \citep{Vigano2013}. The $\kappa_i$ coefficients correspond to different cases for the magnetospheric torque \citep{Philippov}: plasma-filled magnetosphere, $\kappa_0=\kappa_1=1$, with the last coefficient defining whether the alignment is considered ($\kappa_2=1$) or not ($\kappa_2=0$), or classical vacuum dipole ($\kappa_0=$, $\kappa_1=\kappa_2=2/3$).}
\label{tab:models}
\end{table}

\subsection{Radio emission properties}

\begin{figure}
\begin{center}
\includegraphics[width=8cm]{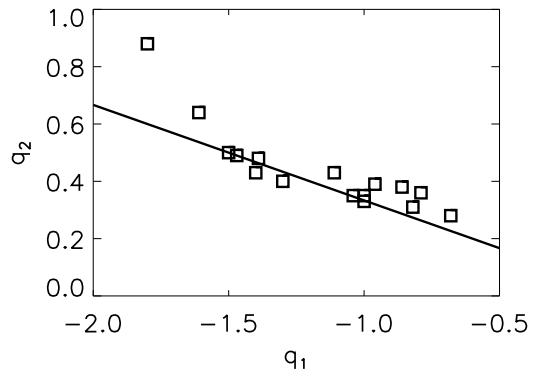}
\caption{Values of the $(q_1,q_2)$ pairs (see eq.~\ref{eq:radiolum}) of the different studies found in the literature. The solid line shows the $q_1=-3 q_2$ condition.}
\label{fig:epepdot}
\end{center}
\end{figure}

The precise physical process that generates the coherent radio beam in pulsars is not yet well understood (e.g., \citealt{melrose95,beskin13}). In lieu of a theoretical model derived from first principles, several authors have analyzed the observed population in order to derive a phenomenological relation between the timing properties ($P$, $\dot{P}$) and the radio-luminosity, defined as $L_{\nu} = d^2 S_{\nu}$ (where $S_{\nu}$ is the detected flux at frequency $\nu$). The usual form is:

\begin{equation}\label{eq:radiolum}
L_{\nu} = L_{0} P^{q_1} \dot{P}^{q_2}~,
\end{equation}
where the values of $L_0, q_1$ and $q_2$ are fitted from observations. Data show a very large dispersion, and the correlations between $L_\nu$ and $P$ or $\dot{P}$ are weak. \cite{ACC} first performed a maximum likelihood analysis of observed pulsars (at $\nu=400$ MHz) and obtained $(q_1,q_2) = (-1.3,0.4)$. Later, numerous alternative fits have been obtained. For example, \cite{Faucher} found $(-1.5,0.5)$ (for $\nu=1400$ MHz). In Fig.~\ref{fig:epepdot} we plot the different proposed pairs of parameters $(q_1,q_2)$, as collected in the recent review by \cite{Lumrev}. The most recent work in this direction \citep{Bates2014} finds that the optimal choices of the exponents are $(q_1,q_2) = (-1.39,0.48)$, very similar to previous studies and to that found for $\gamma-$ray pulsars.

For our purposes, and considering that the computational time increases exponentially with the number of free parameters, we keep the number of free parameters in the radio-luminosity formula as low as possible in order to introduce and explore new parameters related to the physical properties of the NS interiors. Thus, we assume that the radio-luminosity obeys the following equation:

\begin{equation}
L_{\nu} = L_{0} \, 10^{L_{\rm corr}} \, (P^{-3} \dot{P})^{\alpha} \sim \dot{E}_{\rm rot}^{\alpha}~,
\label{radiolum}
\end{equation}
where we fix $L_{0} = 5.69 \times 10^6$ mJy kpc$^2$ and $L_{\rm corr}$ is a random correction chosen from a zero-centered gaussian of $\sigma = 0.8$ (like in \citealt{Faucher}), to take into account the large observed dispersion. The index $\alpha$ is the only free parameter in the luminosity. Varying the value of $\alpha$ corresponds to moving along the solid line in Fig.~\ref{fig:epepdot}, along which most of the proposed values lie.

In order to detect a synthetic pulsar, it must have a flux $S_\nu=L_{\nu}/d^2$ above the threshold $S_{\rm min}$ \citep{Dewey}. For each pulsar, the value of the latter is evaluated considering its value of $P$, the sky-temperature map from \cite{Haslam}, the electron density model from \cite{ne2002} and the appropriate survey parameters \citep{pksmb, swmb}.

Finally, to determine the probability that its radio beam crosses our line of sight, we employ the phenomenological expression obtained by analysing the polarization data for a large number of isolated pulsars \citep{beam}:
\begin{equation}
f(P) = 0.09 \left(\log{\frac{P}{10\; {\rm s}}}\right)^2 +0.03~.
\label{beam}
\end{equation}

\subsection{Procedure to search for optimal parameters}

Each simulation is characterized by the following set of parameters:

\begin{itemize}
\item initial magnetic field distribution: $\mu_{B_0}$, $\sigma_{B_0}$;
\item initial period distribution: $\mu_{P_0}$, $\sigma_{P_0}$;
\item magneto-thermal evolution model (A, B, or C);
\item magnetospheric model (vacuum or plasma-filled magnetospheres);
\item radio-luminosity parameter $\alpha$;
\item initial mean Galactic height $\langle z_{0} \rangle$;
\item initial mean kick velocity $\langle v_{0} \rangle$.
\end{itemize}
Since our computational resources are limited, we cannot explore completely all dimensions. In the remainder of this subsection we describe the algorithm to determine the optimal parameters for each magneto-thermal evolution and magnetospheric model, and the particular approach used to fix $\langle z_{0} \rangle$ and $\langle v_{0} \rangle$. In the next section we will study in detail the sensitivity of the results to each parameter.

\subsubsection{The simulated annealing method}

Our aim is to find the optimal set of parameters that minimizes the differences between the distributions in the synthetic and observed samples. These differences are quantified by the distance $D$ of the 2-D KS test (see Appendix \ref{sec:KS} for definitions and details). We apply the 2-D KS test to $P$ and $\dot{P}$ distributions of the PMB survey, as these are the main observables.

We implemented a numerical algorithm based on the simulated annealing method \citep{numericalrecipes}, which is particularly suited for multi-parameter optimization problems with a global minimum surrounded by local minima. It is based on random walks that search for the minimum energy state, in analogy with the manner liquids cool down, freeze and become crystals. A parameter $T$, playing the role of the temperature in crystallizing systems, is used to control the step of the random walk between different states (i.e., points in the parameter space).

The distance $D$ is the function to be minimized and $\Delta D\equiv D_{\rm new}-D_{\rm old}$ is the difference between the distances computed with the sets of parameters after and before the step, respectively. If $\Delta D < 0$, the change is always accepted, otherwise we only accept the transition with the probability
\begin{equation}
p(D) = \exp{\left(-\frac{\Delta D}{T}\right)}~.
\end{equation}
This is the principal difference with other optimization algorithms that directly go downhill (i.e., the new state is accepted if and only if $\Delta D < 0$). The control parameter $T$ must be tuned to allow enough uphill paths to be able to escape from local minima.

The steps of the algorithm can be summarized as follows.

\begin{enumerate}

\item Define and discretize the space of parameters.
\item Set initial points in the N-dimensional parameter space and calculate
the mean $D$-value after $n_{\rm run}$ realizations ($\bar{D}_{\rm old}$). Typically we use $n_{\rm run} = 10$.
\item Randomly move to a new neighbor point of the discrete grid and calculate $\bar{D}_{\rm new}$.
\item If $\Delta \bar{D} < 0$,  or $\Delta \bar{D} > 0$ and $\rm{Ran}[0,1] < \exp{(-\Delta \bar{D}/T)}$,
move to the new point.  Otherwise a new random step from the old point is taken (go to iii).
\item Repeat this cycle of random steps at fixed $T$ until  a given number of successful steps  
$n_{\rm suc}$ or a maximum number of trials $n_{\rm over}$ is reached. The typical values used in our runs are
$n_{\rm over} \approx 100$ and $n_{\rm suc}=n_{\rm over}/2$.
\item At the end of the process, slightly lower the temperature (typically, by a factor $0.75$), refine the discretization 
grid, and restart the cycle (go to ii).
\item  Continue until the random path has converged to a final state (presumably the minimum).
\end{enumerate}

Convergence is typically reached after about $10-15$ full cycles. The lowest value of D obtained is usually in the range $0.06-0.07$.
For each physical model, we repeated the whole annealing procedure several times and checked that
we reached the same final state even for different initial points in the space of parameters.

\subsubsection{Mean Galactic height and mean velocity distributions.}


\begin{figure}
\begin{center}
\includegraphics[width=8cm]{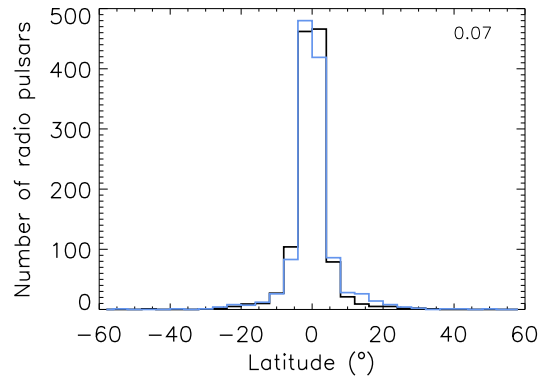}
\end{center}
\caption{Comparison of the observed (PMB+SMB+PHL) and synthetic distributions (model B0) in Galactic latitude with the parameters $\langle z_{0} \rangle =0.18$  kpc and $\langle v_{0} \rangle =380$ km s$^{-1}$. The p-value of the KS test is $0.07$.}
\label{fig:dist_z0v0}
\end{figure}


Since the PMB survey covers a thin region close to the Galactic plane, the synthetic modeling of the PMB sample 
turns out to be rather insensitive to $\langle z_{0} \rangle$, which is also correlated to the parameter 
$\langle v_{0} \rangle$. However, it affects the distribution of pulsars in Galactic latitude, thus changing the ratio of the number of detected pulsars between  SMB and PMB.


For all models (see Table \ref{tab:models}) we find that the best fits more or less overlap in the region delimited by $\langle z_{0} \rangle =[0.05,0.20]$ kpc and $\langle v_{0} \rangle =[300,800]$ km/s. As an example, in Fig. \ref{fig:dist_z0v0} we show the comparison of the observed (PMB, SMB and PHL) and synthetic distributions in Galactic latitude with the parameters $\langle z_{0} \rangle =0.18$  kpc and $\langle v_{0} \rangle =380$ km/s. The p-value of the KS test applied to the two distributions is 0.07. These two parameters have little influence for our purposes in the rest of the paper. Hereafter we fix $\langle v_{0} \rangle=380$ km s$^{-1}$ and $\langle z_{0} \rangle=0.18$ kpc, which ensures that the relative ratio of SMB to PMB sources always stays within reasonable values around 0.2, and that the 
vertical distribution of pulsars is reasonably well reproduced for all models. The ratio of PHL to PMB pulsars ($\sim 0.03$) does not depend strongly on $\langle z_{0} \rangle$ and $\langle v_{0} \rangle$.

\begin{table*}
\begin{center}
\begin{tabular}{|l|cccccccr}
\hline
\hline
Model       & $\mu_{B_0}$ 	& $\sigma_{B_0}$ 	& $\mu_{P_0}$  	&  $\sigma_{P_0}$ & $\alpha$ & $\bar{D}$ & $n_{\rm br}$ & $N_{\rm smb}/N_{\rm pmb}$  \\
       & $\log{B}$ [G] 	& $\log{B}$ [G] & [s]  	&  [s] & & & [century$^{-1}$]   &  \\
\hline
A0	&	$12.65$ 	& $0.50$ 		& $0.38$ 		& $0.35$ 		& $0.50$ & $0.072 \pm 0.008$ & $5.09 \pm 0.15$ & $0.211 \pm 0.014$ \\
B0	&	$13.04$ 	& $0.55$ 		& $0.23$ 		& $0.32$ 		& $0.44$ & $0.068 \pm 0.006$ & $2.20 \pm 0.07$ & $0.184 \pm 0.013$ \\
B0$^\dagger$    &	$13.04$ 	& $0.52$ 		& $0.19$ 		& $0.19$ 		& $0.46$ & $0.076 \pm 0.008$ & $2.00 \pm 0.06$ & $0.181 \pm 0.014$ \\
C0	&	$13.20$ 	& $0.72$ 		& $0.37$ 		& $0.33$ 		& $0.41$ & $0.085 \pm 0.009$ & $3.17 \pm 0.09$ & $0.170 \pm 0.013$  \\
B1	&	$13.20$ 	& $0.62$ 		& $0.22$ 		& $0.42$ 		& $0.44$ & $0.078 \pm 0.009$ & $2.50 \pm 0.08$ & $0.182 \pm 0.014$\\
B2	&	$13.98$ 	& $0.85$ 		& $0.23$ 		& $0.19$ 		& $0.44$ & $0.082 \pm 0.009$ & $3.44 \pm 0.12$ & $0.177 \pm 0.016$ \\
\hline
\hline
\end{tabular}
\end{center}
\caption{Optimal parameter values for the models presented in Table~\ref{tab:models}, and associated $D$-value, birth rate and ratio between pulsars detectable by SMB and PMB. Mean values and statistical dispersions have been estimated by performing $50$ realizations for each model. Model B0$^\dagger$ has the same initial setup as model B0, except the different values of mean Galactic scale, $\langle z_{0} \rangle = 0.10$ kpc, and mean velocity, $\langle v_{0} \rangle = 600$ km s$^{-1}$.}
\label{tab:optimal}
\end{table*}

\begin{figure*}
\begin{center}
\includegraphics[width=8cm]{Images/ppdot_pkssw.eps}
\includegraphics[width=8cm]{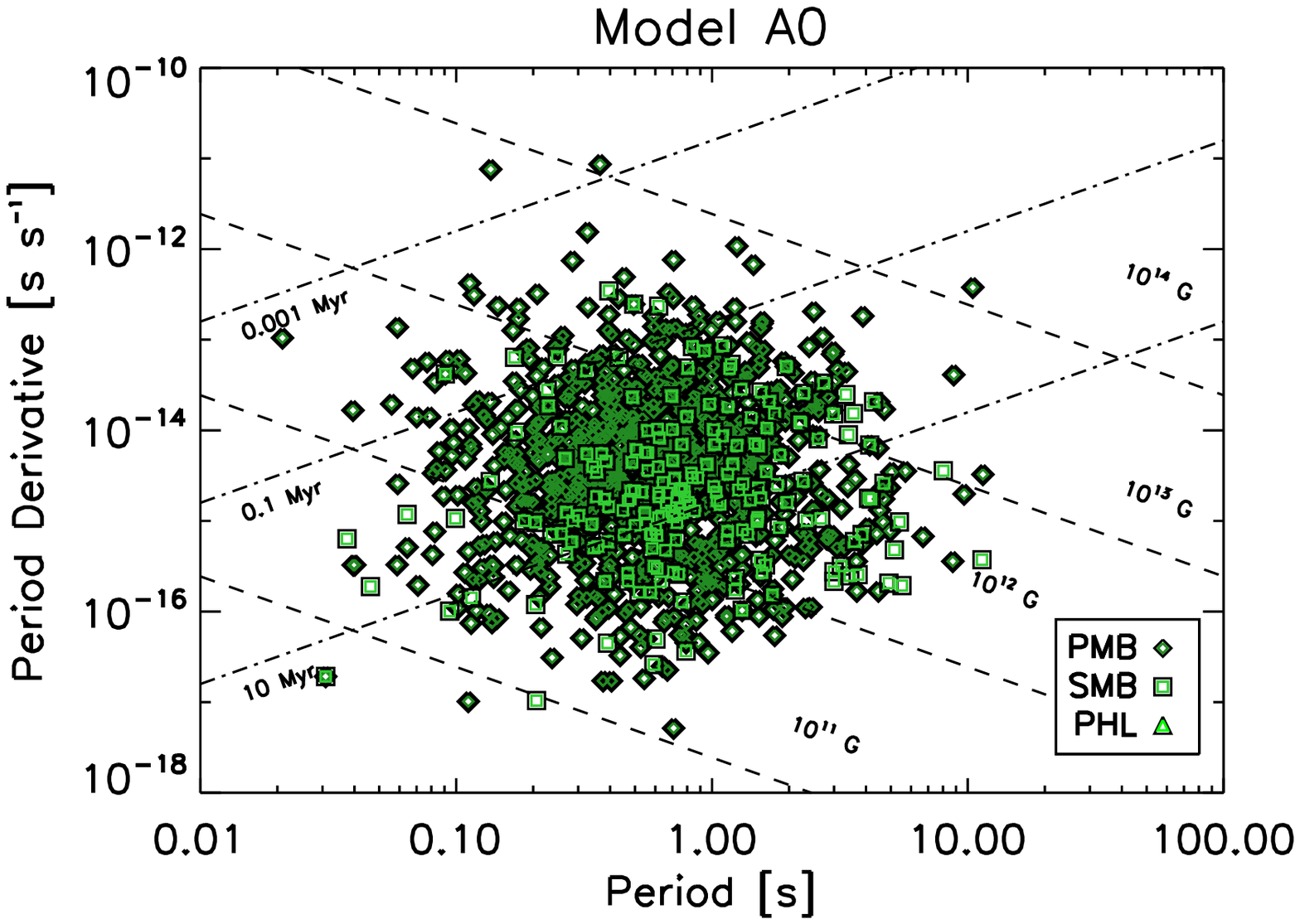}
\includegraphics[width=8cm]{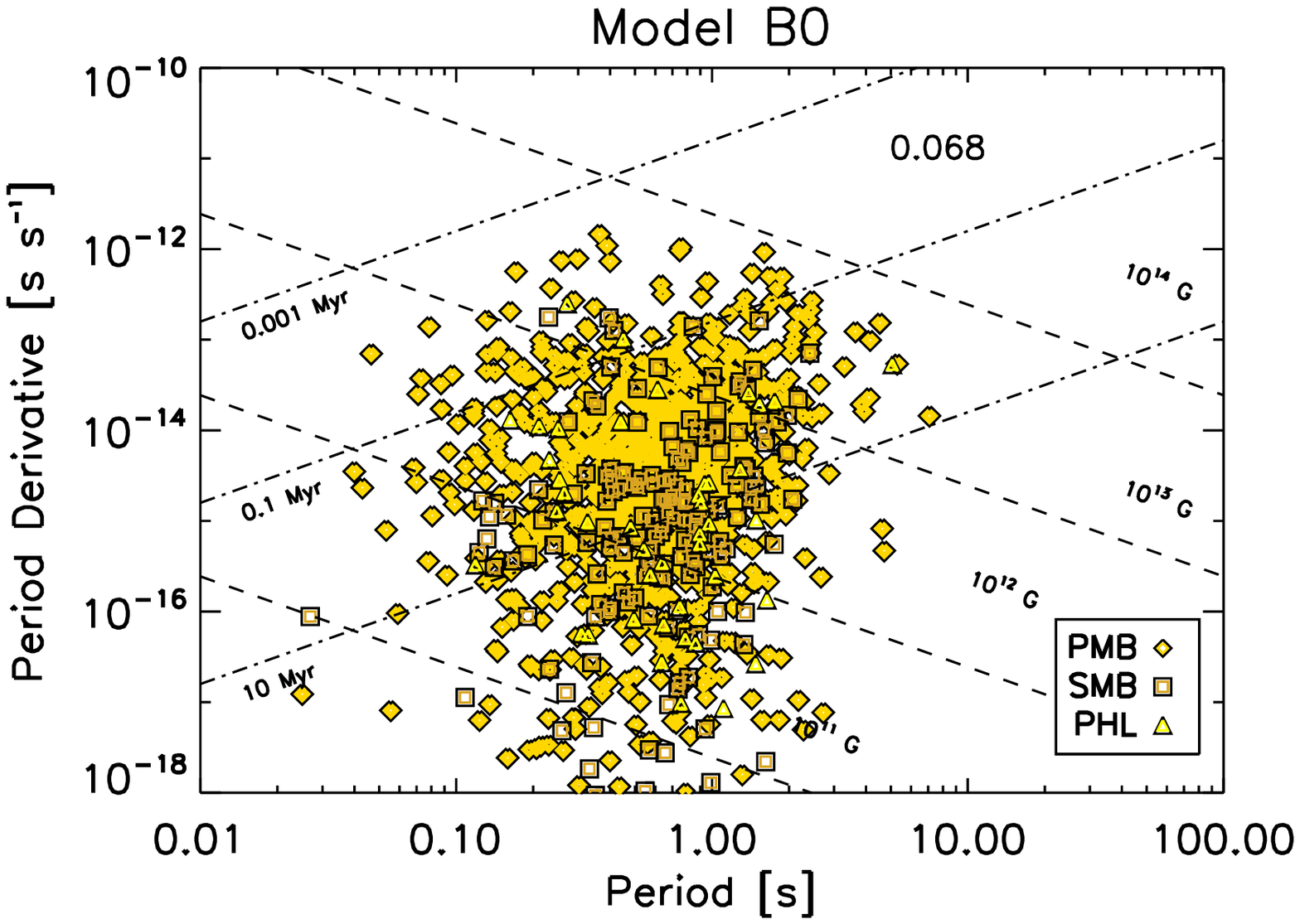}
\includegraphics[width=8cm]{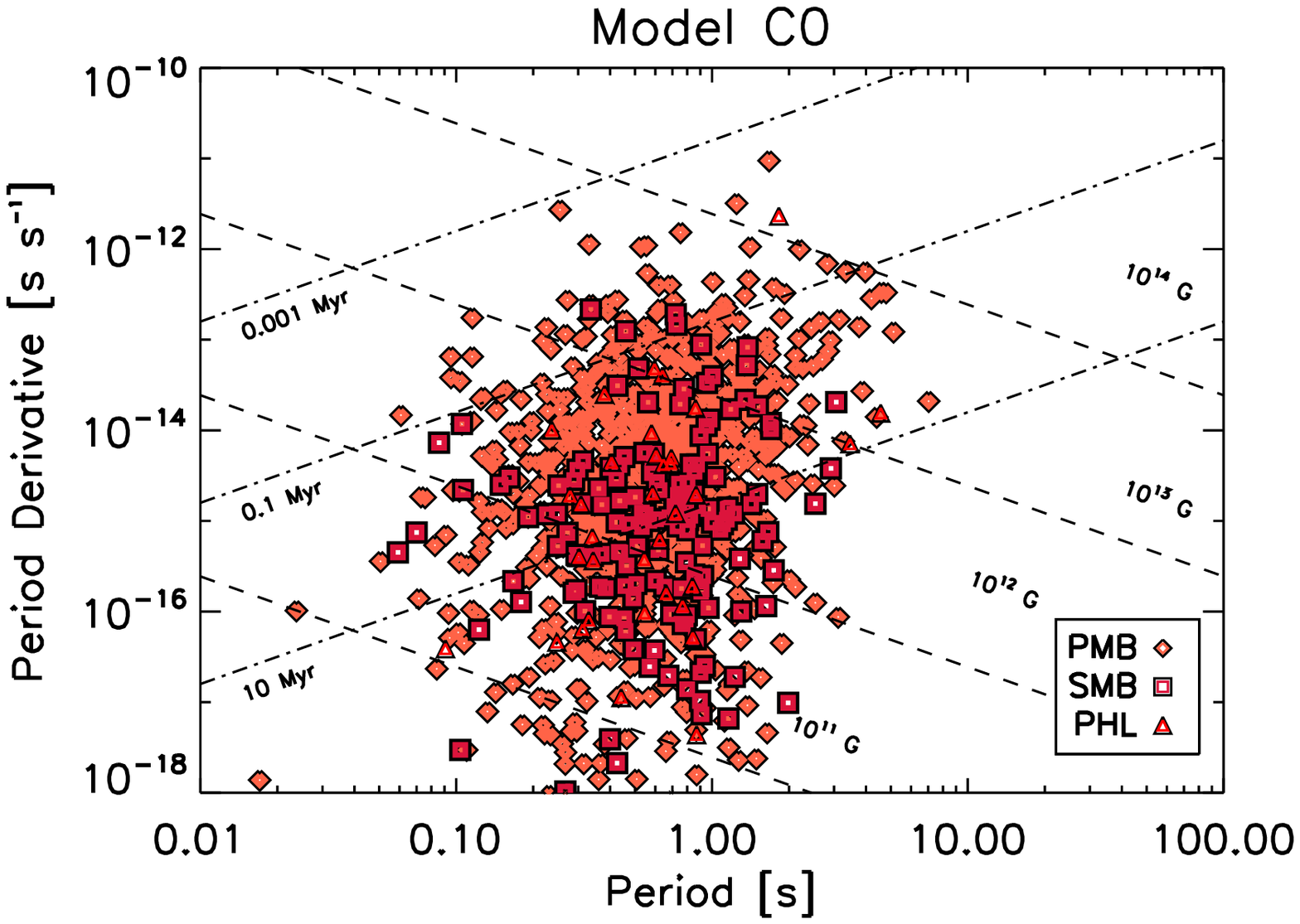}
\includegraphics[width=8cm]{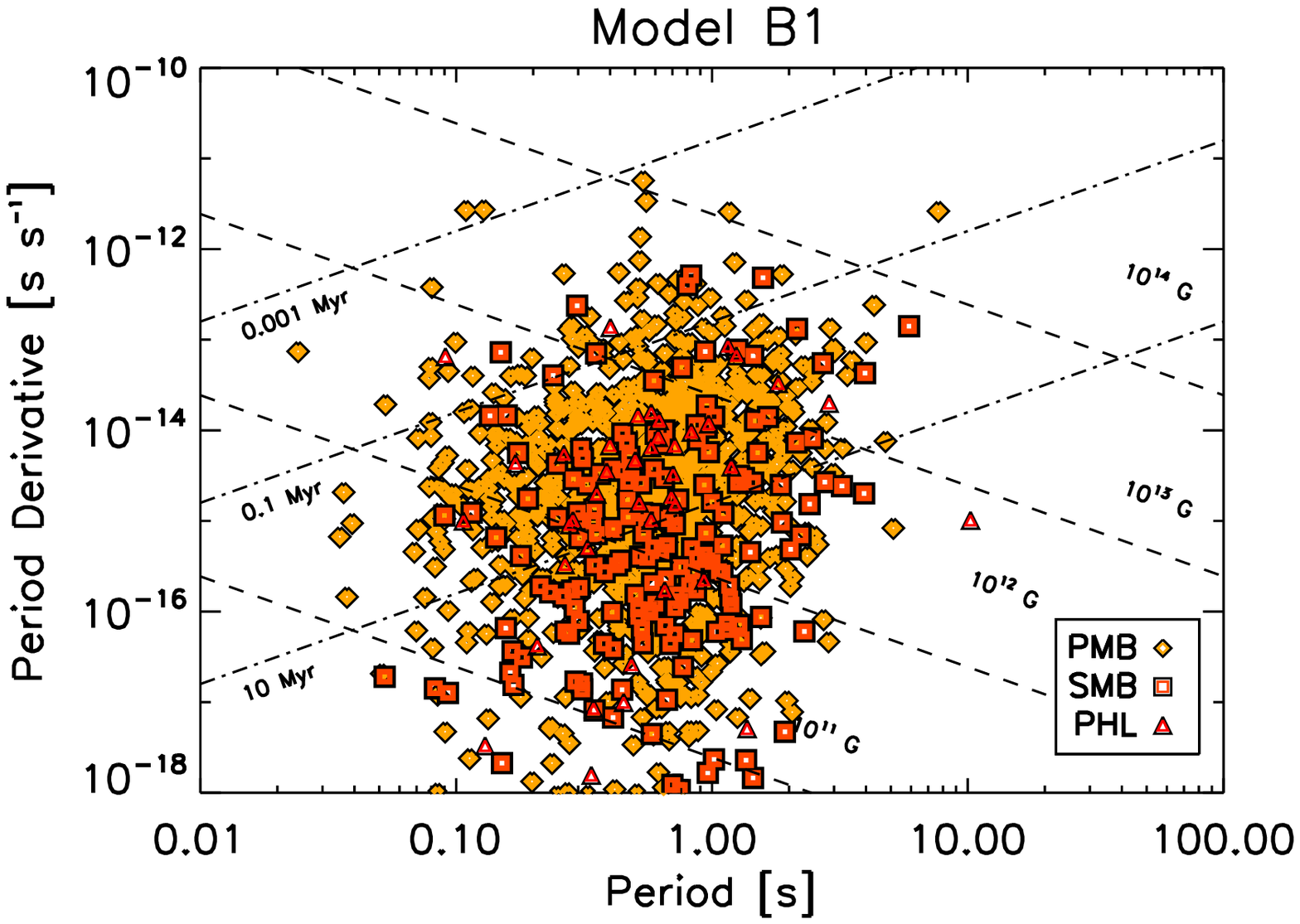}
\includegraphics[width=8cm]{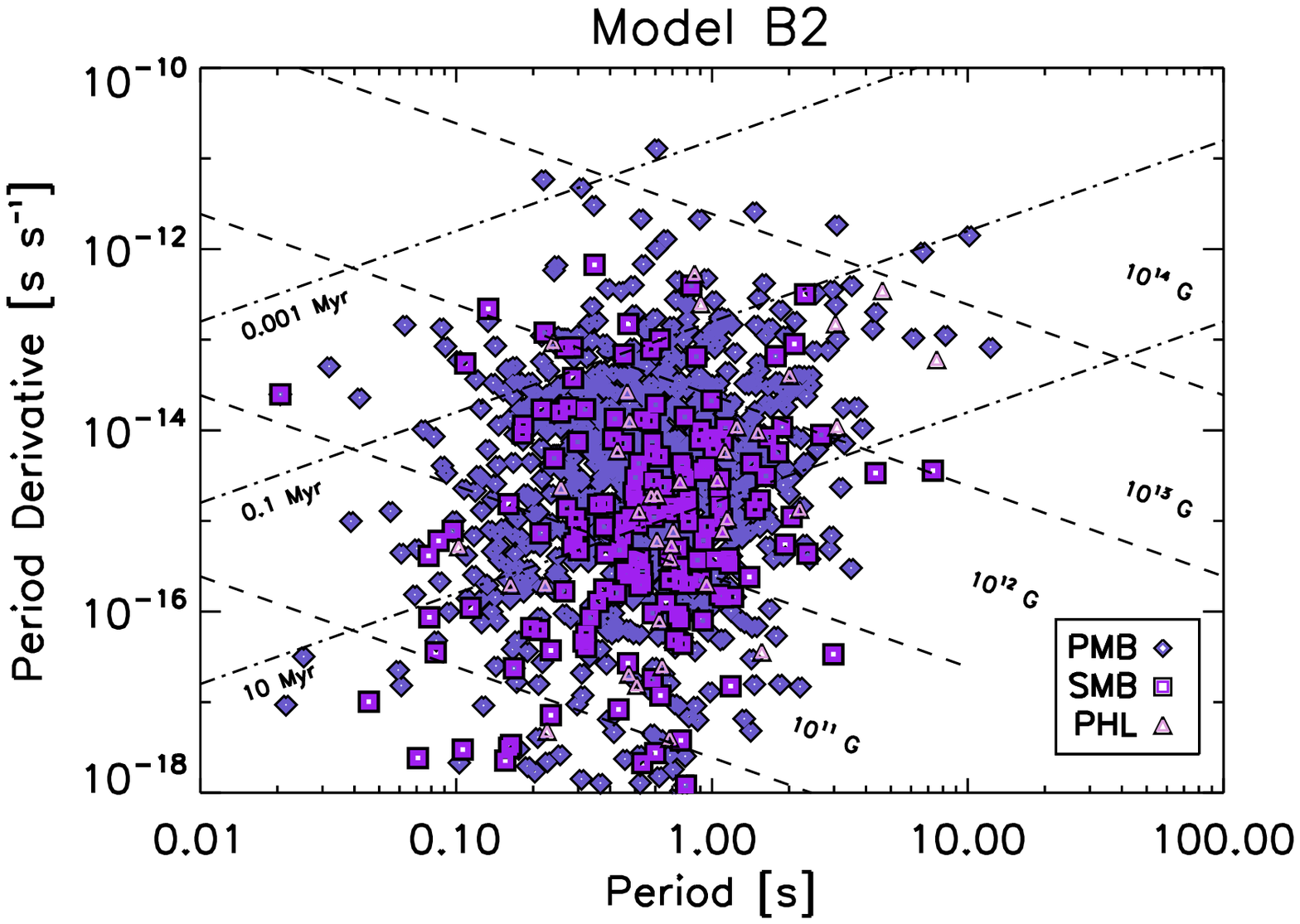}
\end{center}
\caption{$P$-$\dot{P}$ diagrams for the observed population of radio pulsars and for the different models considered in Table~\ref{tab:models}, set up with the corresponding optimal parameters, which are listed in Table~\ref{tab:optimal}.
The graphics correspond to the single realization among the 50.}
\label{fig:ppdot_magnetothermal}
\end{figure*}

\section{Discussion}\label{sec:disc}

We begin our discussion and the exploration of the effects of each relevant parameter by defining the {\it optimal set of parameters} for each NS evolution model. These optimal solutions have been found by performing the annealing procedure with $\mu_{P_0}$, $\mu_{B_0}$, $\sigma_{P_0}$, $\sigma_{B_0}$, and $\alpha$.

Table~\ref{tab:optimal} lists the parameter sets for each one of the five models presented in Table~\ref{tab:models}. Examples of the resulting $P$-$\dot{P}$ diagrams for particular realizations are plotted in Fig.~\ref{fig:ppdot_magnetothermal}. For all models, the mean $D$-value of $50$ realizations is similar (0.06-0.07), and the errors quoted are due to statistical fluctuations between different realizations. The worst results are obtained for model B2, which uses the angle evolution in vacuum, but this is not surprising. Actually no acceptable solution can be found assuming a uniform distribution of the angle $\chi$ at birth, since most pulsars align too fast and stop spinning down, as discussed in the literature \citep{alignment}. The only way we found acceptable solutions for model B2 is to assume a uniform $\chi_0$ distribution in a restricted range $[\chi_{\rm 0,min},90^\circ]$, because the alignment timescale for this model is strongly dependent on the initial angle. This is the result quoted in the table, obtained for $\chi_{\rm 0,min} = 84^\circ$, which is very difficult to justify because there is neither observational evidence nor theoretical argument favoring NSs to be born as nearly orthogonal rotators. For the plasma-filled magnetosphere model (B1), the alignment is slower and, more importantly, the torque is not cancelled even with complete alignment. Thus it is not difficult to find good fits with the initial angle distribution of eq.~(\ref{eq:initial_angle}).

Our results may appear to be contradicting the conclusions of previous works \citep{Ridley2010}, who concluded that magnetic dipole spin-down laws provided a good description of the population without alignment. The origin of the apparent discrepancy stems from the different treatment of alignment in \cite{Ridley2010}. They assume a phenomenological alignment model of the form $\sin\chi = \sin\chi_0 \exp{(-t/t_d)}$ for both the vacuum model and the magnetospheric torque model of \cite{CS2006}. However, one must consistently solve the system of equations \ref{eq:alignment} and \ref{eq:spindown} to obtain the angle evolution. In the vacuum limit, the decay of $\sin\chi$ is actually exponential, but the decay time $t_d$ is not a free parameter, and it turns out to be too fast. This model cannot be reconciled with the observations unless all NSs are born as nearly orthogonal rotators, as discussed above. Conversely, the consistent angle evolution of a magnetospheric model with $\kappa_0 > 0$ is not an exponential decay, but a slower (close to a power-law) decaying function. \cite{Ridley2010} proposed a way to resolve the discrepancy,  by removing the dependence of the spin-down law on $\chi$. Indeed, this is exactly what happens when the consistent approach (solving the system \ref{eq:alignment} and \ref{eq:spindown}) is adopted: since $\kappa_0 \approx \kappa_1$, as $\chi$ becomes small, the evolution of $\dot{P}$ becomes independent of the angle.

\begin{figure}
\begin{center}
\includegraphics[width=8cm]{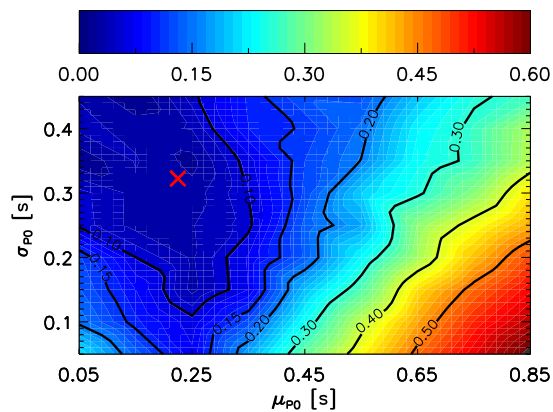}
\end{center}
\caption{Contour plot for the $D$-value in the $\mu_{P_0}$-$\sigma_{P_0}$ plane
for the optimal solution of model B0. The red cross corresponds to the reference value.}
\label{fig:p0}
\end{figure}

The $D$-value is mostly determined by the central region of the $P$-$\dot{P}$ diagram, where the bulk of radio-pulsars are located, and the tails have little weight in the KS test.  We note that the more elongated shapes of models with field decay (B,C) seem to better reproduce the real distribution. In particular, model A has a lack of sources with $\dot{P} < 5 \times 10^{-16}$. We also note that the low-$\dot{P}$ part of the diagram is subject to selection effects, since to measure a low value of $\dot{P}$ requires a very long time span. Such an effect is not included in our PS code, thus our selection procedure, in which we assumed that all pulsars with $\dot{P} > 10^{-18}$ are potentially visible, results in the overestimation of the number of detected low-$\dot{P}$ sources. Thus we conclude that models B and C are favored compared to model A.

Another important issue is the NS birth rate, $n_{\rm br}$, eq.~(\ref{eq:birthrate}), since it could be used to partially 
remove the degeneracy. The typical values of the birth rate are in the range 1-3 NS per century, estimated in different
ways, from pulsar current analysis, to population synthesis \citep{Vranesevic04,Lorimer06,Faucher}. 
In our study, the optimal parameterizations for models with field decay predict a birth rate of 2-3 per century, while for model A it is somewhat higher (5 per century). Although this value is still acceptable given the many existing uncertainties, it seems to be higher than the usual expectations, which again disfavors this model with respect to the others.

We note that $n_{\rm br}$ is higher for models with larger $\mu_{B_{0}}$ and $\mu_{P_{0}}$, for which pulsars 
tend to have longer periods and narrower radio beams.
In addition, when $\alpha$ is reduced the contours of constant radio-luminosity are shifted to the right in the $P$ - $\dot{P}$ diagram, increasing the detectability of old stars and thus decreasing the birth rate. Finally, we note that the kinematic parameters $\langle z_{0} \rangle$ and $\langle v_{0} \rangle$ affect $n_{\rm br}$ to a lesser extent: as we increase one (or both) of them, more NSs are able to escape from the Galactic plane and the birth rate needs to be slightly higher to compensate.


\begin{figure}
\begin{center}
\includegraphics[width=8cm]{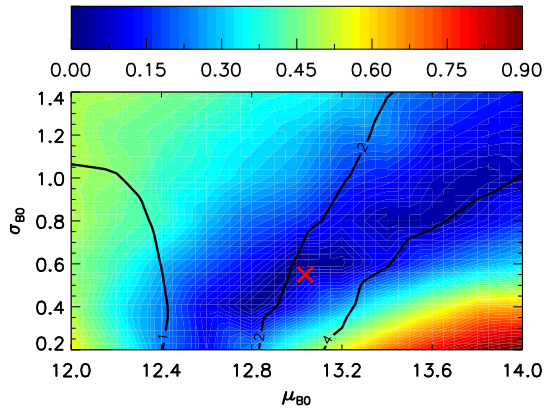}
\end{center}
\caption{Contour plot for the $D$-value in the $\mu_{B_0}$ - $\sigma_{B_0}$ plane
for optimal solution of model B0.
The red cross corresponds to the reference value.
The solid lines are the contours of constant birth rate of 1,2, and 4 NSs per century.}
\label{fig:b0}
\end{figure}

\subsection{Initial period distribution}

We turn now to explore the influence of the initial distribution of spin periods in the 
observed population. In Fig.  \ref{fig:p0} we show $D$-value contours in the plane $\mu_{P_0}$-$\sigma_{P_0}$,
keeping fixed the rest of parameters corresponding to the optimal solution of model B0 (see Table 
\ref{tab:optimal}). For the Gaussian distribution we used, we found a large region with acceptable
low values of $D$, basically any distribution with $\mu_{P_0}<0.4$ s and $\sigma_{P_0}>0.2$ s may fit the data.
We also tried a uniform distribution with an upper cut-off $P_{\rm max}$, and found acceptable fits for values of 
$P_{\rm max} \approx 0.5$ s. The main conclusion, already pointed out by \cite{Gon2004}, is that the
observed population is quite insensitive to the initial period distributions. Practically any nearly uniform distribution in the
range $0<P_0	\lesssim 0.5$ s can fit the data by slightly readjusting the rest of parameters.

\subsection{Initial magnetic field distribution}

Constraining the initial magnetic field distribution is one of the most important objectives of population synthesis
studies. From the results in Table \ref{tab:optimal}, we can see that models with strong field decay require an initial
field distribution centered in a value a factor of 3 higher than models without field evolution.
To study how degenerate this result is, we have proceeded as for the period, and varied 
$\mu_{B_0}$ and  $\sigma_{B_0}$,
keeping fixed the rest of parameters corresponding to model B0. The results for the
$D$-value contours are shown in Fig.~\ref{fig:b0}. Contrary to the period distribution, here we observe a strong
correlation between the two parameters. There is a long valley with acceptable solutions, with similar statistical 
significance to the ``optimal" solution (red cross). 
The $D$-value differences between near regions in the valley is lower than the statistical fluctuations between different
realizations, and a broader distribution centered in higher values of the magnetic field can also reproduce 
the same properties of the radio pulsar distribution. Similar results are obtained with the other models, but note that the 
degeneracy does not extend to arbitrary low values of $\mu_{B_0}$: for the models with field decay, $\mu_{B_0} > 12.7$ 
seems to be required. We also overplot the contours corresponding to a constant NS birth rate of 1, 2 and 4 per century to show that an independent estimate of this value could break the degeneracy.

\subsection{Radio-luminosity parameter $\alpha$}

\begin{figure}
\begin{center}
\includegraphics[width=8cm]{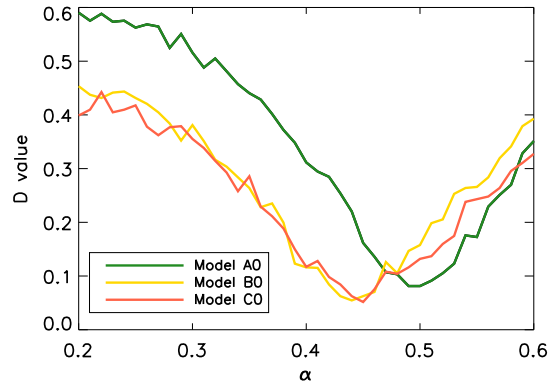}
\end{center}
\caption{$D$-value as a function of $\alpha$ for models A0, B0 and C0.}
\label{fig:d_alpha}
\end{figure}

The radio-luminosity power index $\alpha$ is more constrained than other parameters.
In Fig. \ref{fig:d_alpha} we have plotted the dependence of the $D$-value with $\alpha$
for the three magneto-thermal evolution models. All of them exhibit well defined minima around the 
same range. For model A0, acceptable values are in the range $[0.45, 0.50]$,
while for models B0 and C0 the minima are found in the region $ [0.42, 0.46]$. The general trend is that a
lower value of $\alpha$ is preferred for models with stronger field decay.
If we look again at Fig.~\ref{fig:epepdot}, the large spread in the values of the coefficients is severely
constrained. Our results only allow a narrow region in the range $p \approx [0.4,0.5]$,
$q \approx [-1.2,-1.5]$. It should be noted, however, that we obtain this constraint by fitting surveys
at 1400 MHz, and the relation may not hold at other frequencies.

A last important remark concerns the interpretation of the so-called death line. We note that, contrary to other authors,
we have not imposed any death line condition for the detectability of radio pulsars (i.e., an abrupt switch off of the radio mechanism). Instead, 
the assumption $L_{\nu} \sim \dot{E}_{\rm rot}^{\alpha}$, combined with observational selection effects, naturally 
accounts for the fading radio emission of pulsars with large $P$ and low $\dot{P}$.
This is in contrast with the classical view in which pulsars simply ``switch-off" when they cross the death line, for example, 
because the electrical voltage drops below a critical value.
\cite{Szary2014} propose an alternative formulation for the death line by imposing the switch-off condition on 
the radio-efficiency,  defined as $\epsilon=L/\dot{E}_{\rm rot}$, and assuming $L$ to be independent of $\dot{E}_{\rm rot}$ 
($\alpha=0$ in eq. (9), justified by the weak observational correlations). Then, by definition, $\epsilon$ anti-correlates 
with $\dot{E}_{\rm rot}$, and one can introduce an equivalent death line ($\epsilon_{\rm max}=0.01$ in their work) above which 
the radio emission is assumed to stop.
In both approaches, a few pulsars can still be detectable in the graveyard under favorable conditions (small distance, 
lucky viewing angle), as seen both in our simulations and in real observations, namely for the very close pulsar PSR J2144-3933, 
with $P=8.5$ s and $\dot{P}\sim 5\times 10^{-16}$ \citep{young99}.

\section{Summary}\label{sec:conc}

In this paper, we presented population synthesis calculations of observable radio pulsars
including the effects of the coupled magnetic, rotational and thermal evolution.
We explored the sensitivity of the results to the initial distribution of period and magnetic field, 
the spatial distribution of NSs at birth, and the radio-luminosity power, with special attention to show
and explain all the different degeneracies in the large parameter space. 

The PMB survey, due to the limited coverage of small Galactic latitudes is not very sensitive to the spatial 
distribution of NS at birth, but the combined use of PMB, SMB and their extensions \citep{Burgay2006,Jacoby2009} 
allows one to constrain the initial velocity and vertical spatial distributions. 
However, there is a strong correlation between the two, and an independent
determination of one of them, or more statistics, are required to give more accurate constraints. 
We find reasonable agreement with the observations for values close to $\langle v_{0} \rangle \approx 300-400$ 
km s$^{-1}$ and   $\langle z_{0} \rangle \approx 0.15-0.2$ kpc.
Having fixed these two parameters, we explored in detail the dependences on the rest of the free parameters. 
Assuming that the radio-luminosity depends on the rotation power in the form of a power-law (with strong dithering),
our results are consistent with $\alpha$ in  in the range $0.45-0.5$. We have not found any strong dependence of
our results on the particular magnetosphere/alignment model, except in the case in which we combine alignment
with vacuum, which is strongly disfavored unless all NSs are born as nearly orthogonal rotators. The large parameter
space allows one to find acceptable solutions for models with and without field decay, with the consistent alignment
angle evolution. For the particular value of the luminosity coefficient ($L_0$) employed,  the model without magnetic field
predicts a relatively large birth rate, and would be disfavored. The caveat is that the birth rate could be 
renormalized by leaving $L_0$ as a free parameter \citep{Gon2004,Gon2007}.

The most interesting results concern the initial distributions of period and magnetic field.
We find that the differences in the initial period distribution between different models are not significant. Indeed, 
any distribution with nearly constant probability density in the range $0<{P_0}<0.5$ s can fit the data, 
but it is important to note that narrow distributions peaked at short spin 
periods ($P_0<100$ ms) cannot be reconciled with the observations. This implies that NSs born as fast 
rotating pulsars are quite rare, or that there is an additional spin-down mechanism operating for a short time, 
effective enough to quickly bring all NSs to periods of a fraction of a second.

The magnetic field initial distribution is better constrained than the period distribution, but the central value and width of the 
distribution are degenerate and strongly correlated: the larger the central value, the wider the distribution. 
This holds true for models with and without field decay, and with different assumptions for the magnetospheric torque.
The optimal value for the fiducial model B1 is about $\log{\frac{B_0}{\rm{G}}} \approx 13.2$, $\sigma_{B_0}\approx 0.6$,
very similar to the result of \cite{PSB_Popov}.
Other models, in particular the model without field decay A0, can also reproduce the observed distributions. However, 
the model without field decay also requires a rather high birth rate (5 NSs per century) while the optimal solutions
for the rest of models correspond to a birth rate of 2-3 pulsars per century. Reducing $L_0$ by about a factor of 100 will
bring back the birth rate of the models with and without field decay to about 1 and 2 pulsars per century, respectively.
Therefore, an independent estimate of the birth rate combined with improvements in the theoretical luminosity function
may be used to break the degeneracy and to discern between different models. If the assumed value of
$L_0= 5.69 \times 10^6$ mJy kpc$^2$ is correct within an order of magnitude, models with field decay are favored against the 
model without field evolution.
Another way to break the degeneracy is to correlate the information obtained by fitting the radio pulsar population
with similar studies in the X-ray and $\gamma-$ray band, a work in progress that will be reported in a separate paper.

\section*{Acknowledgments}

This work was supported in part by the grants AYA 2010-21097-C03-02 and Prometeo/2009/103, and by 
the {\it New Compstar} COST action MP1304. MG is supported by the fellowship BES-2011-049123. DV is supported by the grants AYA2012-39303 and SGR2009-811. We gratefully acknowledge useful discussions and suggestions from Nanda Rea and Peter Gonthier.

\appendix

\section{Kolmogorov-Smirnov goodness-of-fit test} \label{sec:KS}

\begin{figure}
\begin{center}
\includegraphics[width=8cm]{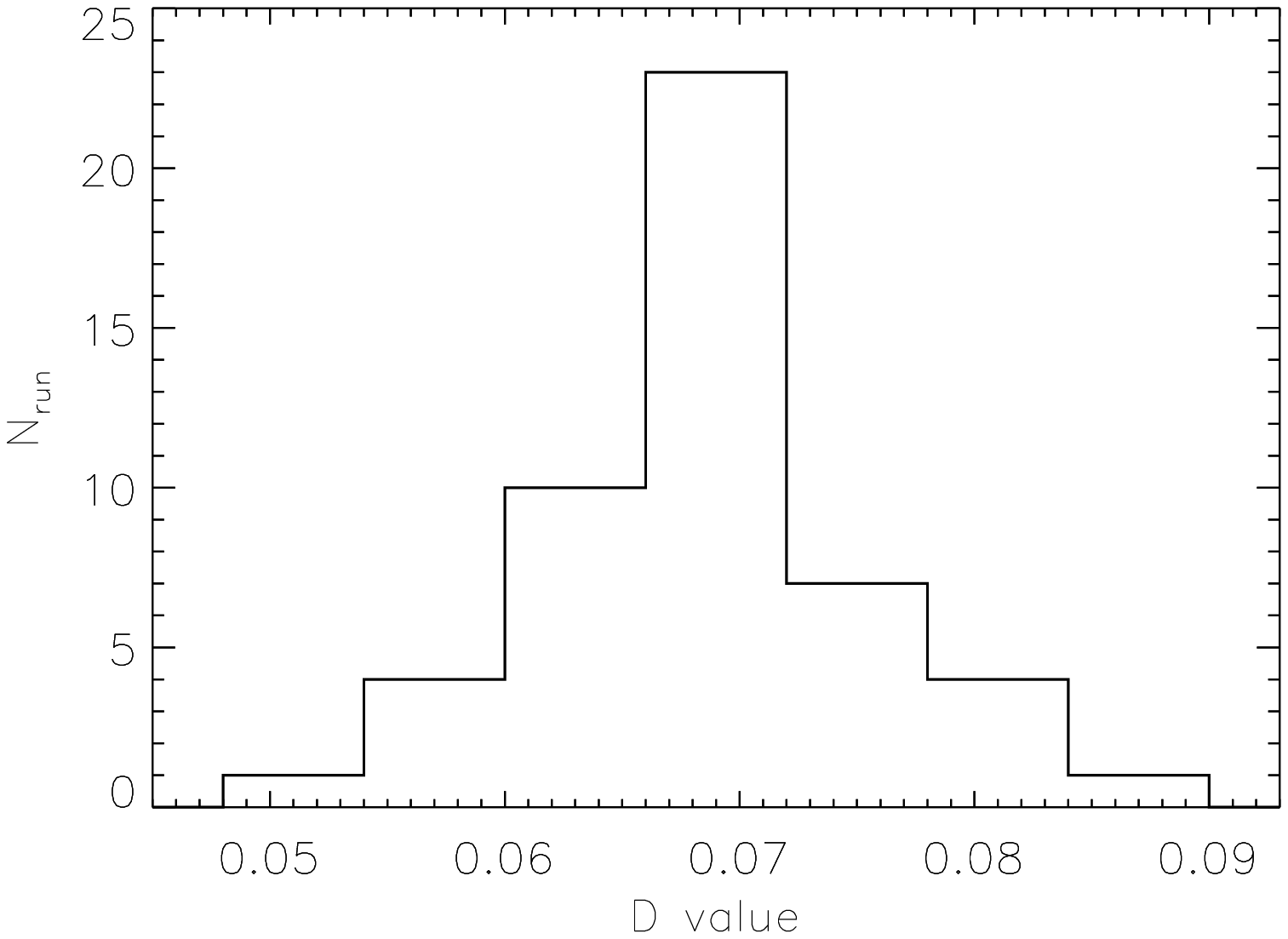}
\end{center}
\caption{Histogram for the different $D$-values obtained in different realizations with the same set of parameters.}
\label{fig:dvalues1}
\end{figure}

The KS test is a well-known statistical method
used to test if the origin of two samples is the same statistical distribution (null hypothesis).
In its one-dimensional form the statistic $D$ is defined as the maximum distance between the two cumulative distributions
of a single measured variable $x$:

\begin{equation}
 D = \max |S_{\rm obs}(x) - S_{\rm sim}(x)|~,
\end{equation}
where $S_{\rm obs}$ and $S_{\rm sim}$ are the observed and simulated cumulative distributions, respectively. 
A significance value can be associated to $D$ as:

\begin{equation}
p (D > D_{\rm obs}) = Q \left((\sqrt{N_{\rm eff}} + 0.12 + \frac{0.11}{\sqrt{N_{\rm eff}}}) D \right)~,
\end{equation}
where $Q$ is a monotonic function given by $Q (\lambda) = 2 \sum^{\infty}_{j=1} (-1)^{j-1} \exp{-2 j^2 \lambda^2}$
and $N_{\rm eff} = N_1 N_2/(N_1+N_2)$, where $N_i$ is the number of elements on each sample.
This $p$-value can be taken as the probability that the two samples come from the same parent distribution. 

A generalized 2-D KS test, although not so rigorous, can also be used (\citealt{numericalrecipes}).
The statistic is estimated from the four natural quadrants that define each point in the plane.
The fraction of data from both samples is calculated for each quadrant.
The extended $D$-value is then defined as the maximum fraction difference found over all points.
The significance $p$ has a similar expression, but including the Pearson correlation coefficient $r$: 

\begin{equation}
p (D > D_{\rm obs}) = Q \left(\frac{\sqrt{N_{\rm eff}} D}{1 + \sqrt{1 - r^2} (0.25 - 0.75 /\sqrt{N_{\rm eff}})} \right)~.
\end{equation}
The differences between two samples drawn from the same initial parameters are related to the effective number $N_{\rm eff}$: the higher the value of $N_{\rm eff}$, the smaller the $D$-value fluctuations.
In Fig.~\ref{fig:dvalues1} we have plotted the $D$-values for $50$ realizations of the code with the same set of parameters but varying the initial random seed. 
The histogram shape resembles a gaussian distribution, with a standard deviation which is $\Delta D_{\rm stat} \sim 0.01$ in most cases. In order to minimize the effect of such statistical fluctuations, one must average over several realizations for each particular parameter set.

\bibliography{references}

\begin{thebibliography}{41}
\expandafter\ifx\csname natexlab\endcsname\relax\def\natexlab#1{#1}\fi

\bibitem[{{Aguilera}, {Pons} \& {Miralles}(2008){Aguilera}, {Pons}, \&
  {Miralles}}]{Aguilera}
{Aguilera} D.~N., {Pons} J.~A., {Miralles} J.~A., 2008, \aap, 486, 255

\bibitem[{{Arzoumanian}, {Chernoff} \& {Cordes}(2002){Arzoumanian}, {Chernoff},
  \& {Cordes}}]{ACC}
{Arzoumanian} Z., {Chernoff} D.~F., {Cordes} J.~M., 2002, \apj, 568, 289

\bibitem[{{Bagchi}(2013)}]{Lumrev}
{Bagchi} M., 2013, International Journal of Modern Physics D, 22, 30021

\bibitem[{{Bates} {et~al}\mbox{.}(2014){Bates}, {Lorimer}, {Rane}, \&
  {Swiggum}}]{Bates2014}
{Bates} S.~D., {Lorimer} D.~R., {Rane} A., {Swiggum} J., 2014, \mnras, 439,
  2893

\bibitem[{{Beskin}, {Istomin} \& {Philippov}(2013{\natexlab{a}}){Beskin},
  {Istomin}, \& {Philippov}}]{Beskin}
{Beskin} V.~S., {Istomin} Y.~N., {Philippov} A.~A., 2013{\natexlab{a}}, Physics
  Uspekhi, 56, 164

\bibitem[{{Beskin}, {Istomin} \& {Philippov}(2013{\natexlab{b}}){Beskin},
  {Istomin}, \& {Philippov}}]{beskin13}
{Beskin} V.~S., {Istomin} Y.~N., {Philippov} A.~A., 2013{\natexlab{b}}, Physics
  Uspekhi, 56, 164

\bibitem[{{Burgay} {et~al}\mbox{.}(2006){Burgay}, {Joshi}, {D'Amico},
  {Possenti}, {Lyne}, {Manchester}, {McLaughlin}, {Kramer}, {Camilo}, \&
  {Freire}}]{Burgay2006}
{Burgay} M. {et~al.}, 2006, \mnras, 368, 283

\bibitem[{{Carlberg} \& {Innanen}(1987)}]{galpot2}
{Carlberg} R.~G., {Innanen} K.~A., 1987, \aj, 94, 666

\bibitem[{{Contopoulos} \& {Spitkovsky}(2006)}]{CS2006}
{Contopoulos} I., {Spitkovsky} A., 2006, \apj, 643, 1139

\bibitem[{{Cordes} \& {Lazio}(2002)}]{ne2002}
{Cordes} J.~M., {Lazio} T.~J.~W., 2002, arXiv:astro-ph/0207156

\bibitem[{{Dewey} {et~al}\mbox{.}(1985){Dewey}, {Taylor}, {Weisberg}, \&
  {Stokes}}]{Dewey}
{Dewey} R.~J., {Taylor} J.~H., {Weisberg} J.~M., {Stokes} G.~H., 1985, \apj,
  294, L25

\bibitem[{{Edwards} {et~al}\mbox{.}(2001){Edwards}, {Bailes}, {van Straten}, \&
  {Britton}}]{swmb}
{Edwards} R.~T., {Bailes} M., {van Straten} W., {Britton} M.~C., 2001, \mnras,
  326, 358

\bibitem[{{Faucher-Gigu{\`e}re} \& {Kaspi}(2006)}]{Faucher}
{Faucher-Gigu{\`e}re} C.-A., {Kaspi} V.~M., 2006, \apj, 643, 332

\bibitem[{{Gonthier} {et~al}\mbox{.}(2007){Gonthier}, {Story}, {Clow}, \&
  {Harding}}]{Gon2007}
{Gonthier} P.~L., {Story} S.~A., {Clow} B.~D., {Harding} A.~K., 2007, \apss,
  309, 245

\bibitem[{{Gonthier}, {Van Guilder} \& {Harding}(2004){Gonthier}, {Van
  Guilder}, \& {Harding}}]{Gon2004}
{Gonthier} P.~L., {Van Guilder} R., {Harding} A.~K., 2004, \apj, 604, 775

\bibitem[{{Haslam} {et~al}\mbox{.}(1995){Haslam}, {Salter}, {Stoffel}, \&
  {Wilson}}]{Haslam}
{Haslam} C.~G.~T., {Salter} C.~J., {Stoffel} H., {Wilson} W.~E., 1995,
  Astronomy Data Image Library, 1

\bibitem[{{Jacoby} {et~al}\mbox{.}(2009){Jacoby}, {Bailes}, {Ord}, {Edwards},
  \& {Kulkarni}}]{Jacoby2009}
{Jacoby} B.~A., {Bailes} M., {Ord} S.~M., {Edwards} R.~T., {Kulkarni} S.~R.,
  2009, \apj, 699, 2009

\bibitem[{{Kuijken} \& {Gilmore}(1989)}]{galpot1}
{Kuijken} K., {Gilmore} G., 1989, \mnras, 239, 651

\bibitem[{{Lorimer} {et~al}\mbox{.}(2006){Lorimer}, {Faulkner}, {Lyne},
  {Manchester}, {Kramer}, {McLaughlin}, {Hobbs}, {Possenti}, {Stairs},
  {Camilo}, {Burgay}, {D'Amico}, {Corongiu}, \& {Crawford}}]{Lorimer06}
{Lorimer} D.~R. {et~al.}, 2006, \mnras, 372, 777

\bibitem[{{Lyne} {et~al}\mbox{.}(2013){Lyne}, {Graham-Smith}, {Weltevrede},
  {Jordan}, {Stappers}, {Bassa}, \& {Kramer}}]{Lyne2013}
{Lyne} A., {Graham-Smith} F., {Weltevrede} P., {Jordan} C., {Stappers} B.,
  {Bassa} C., {Kramer} M., 2013, Science, 342, 598

\bibitem[{{Manchester} {et~al}\mbox{.}(2005){Manchester}, {Hobbs}, {Teoh}, \&
  {Hobbs}}]{ATNF}
{Manchester} R.~N., {Hobbs} G.~B., {Teoh} A., {Hobbs} M., 2005, \aj, 129, 1993

\bibitem[{{Manchester} {et~al}\mbox{.}(2001){Manchester}, {Lyne}, {Camilo},
  {Bell}, {Kaspi}, {D'Amico}, {McKay}, {Crawford}, {Stairs}, {Possenti},
  {Kramer}, \& {Sheppard}}]{pksmb}
{Manchester} R.~N. {et~al.}, 2001, \mnras, 328, 17

\bibitem[{{Melrose}(1995)}]{melrose95}
{Melrose} D.~B., 1995, Journal of Astrophysics and Astronomy, 16, 137

\bibitem[{{Michel} \& {Goldwire}(1970)}]{alignment}
{Michel} F.~C., {Goldwire}, Jr. H.~C., 1970, \aplett, 5, 21

\bibitem[{{Philippov}, {Tchekhovskoy} \& {Li}(2013){Philippov}, {Tchekhovskoy},
  \& {Li}}]{Philippov}
{Philippov} A., {Tchekhovskoy} A., {Li} J.~G., 2013, arXiv:astro-ph/1311.1513

\bibitem[{{Pierbattista} {et~al}\mbox{.}(2012){Pierbattista}, {Grenier},
  {Harding}, \& {Gonthier}}]{Pierbattista}
{Pierbattista} M., {Grenier} I.~A., {Harding} A.~K., {Gonthier} P.~L., 2012,
  \aap, 545, A42

\bibitem[{{Pons}, {Miralles} \& {Geppert}(2009){Pons}, {Miralles}, \&
  {Geppert}}]{Pons2009}
{Pons} J.~A., {Miralles} J.~A., {Geppert} U., 2009, \aap, 496, 207

\bibitem[{{Pons}, {Vigan{\`o}} \& {Rea}(2013){Pons}, {Vigan{\`o}}, \&
  {Rea}}]{Pons2013}
{Pons} J.~A., {Vigan{\`o}} D., {Rea} N., 2013, Nature Physics, 9, 431

\bibitem[{{Popov} {et~al}\mbox{.}(2010){Popov}, {Pons}, {Miralles}, {Boldin},
  \& {Posselt}}]{PSB_Popov}
{Popov} S.~B., {Pons} J.~A., {Miralles} J.~A., {Boldin} P.~A., {Posselt} B.,
  2010, \mnras, 401, 2675

\bibitem[{{Popov} \& {Prokhorov}(2007)}]{PSrev}
{Popov} S.~B., {Prokhorov} M.~E., 2007, Physics Uspekhi, 50, 1123

\bibitem[{Press {et~al}\mbox{.}(1993)Press, Teukolsky, Vetterling, \&
  Flannery}]{numericalrecipes}
Press W.~H., Teukolsky S.~A., Vetterling W.~T., Flannery B.~P., 1993, Numerical
  Recipes in FORTRAN; The Art of Scientific Computing, 2nd edn. Cambridge
  University Press, New York, NY, USA

\bibitem[{{Ridley} \& {Lorimer}(2010)}]{Ridley2010}
{Ridley} J.~P., {Lorimer} D.~R., 2010, \mnras, 404, 1081

\bibitem[{{Szary} {et~al}\mbox{.}(2014){Szary}, {Zhang}, {Melikidze}, {Gil}, \&
  {Xu}}]{Szary2014}
{Szary} A., {Zhang} B., {Melikidze} G.~I., {Gil} J., {Xu} R.-X., 2014, \apj,
  784, 59

\bibitem[{{Tauris} \& {Manchester}(1998)}]{beam}
{Tauris} T.~M., {Manchester} R.~N., 1998, \mnras, 298, 625

\bibitem[{{Vigan{\`o}} {et~al}\mbox{.}(2013){Vigan{\`o}}, {Rea}, {Pons},
  {Perna}, {Aguilera}, \& {Miralles}}]{Vigano2013}
{Vigan{\`o}} D., {Rea} N., {Pons} J.~A., {Perna} R., {Aguilera} D.~N.,
  {Miralles} J.~A., 2013, \mnras, 434, 123

\bibitem[{{Vranesevic} {et~al}\mbox{.}(2004){Vranesevic}, {Manchester},
  {Lorimer}, {Hobbs}, {Lyne}, {Kramer}, {Camilo}, {Stairs}, {Kaspi}, {D'Amico},
  {Possenti}, {Crawford}, {Faulkner}, \& {McLaughlin}}]{Vranesevic04}
{Vranesevic} N. {et~al.}, 2004, \apjl, 617, L139

\bibitem[{{Wainscoat} {et~al}\mbox{.}(1992){Wainscoat}, {Cohen}, {Volk},
  {Walker}, \& {Schwartz}}]{Milky_way}
{Wainscoat} R.~J., {Cohen} M., {Volk} K., {Walker} H.~J., {Schwartz} D.~E.,
  1992, \apjs, 83, 111

\bibitem[{{Weltevrede} \& {Johnston}(2008)}]{WJ2008}
{Weltevrede} P., {Johnston} S., 2008, \mnras, 387, 1755

\bibitem[{{Young}, {Manchester} \& {Johnston}(1999){Young}, {Manchester}, \&
  {Johnston}}]{young99}
{Young} M.~D., {Manchester} R.~N., {Johnston} S., 1999, \nat, 400, 848

\bibitem[{{Young} {et~al}\mbox{.}(2010){Young}, {Chan}, {Burman}, \&
  {Blair}}]{Young2010}
{Young} M.~D.~T., {Chan} L.~S., {Burman} R.~R., {Blair} D.~G., 2010, \mnras,
  402, 1317

\bibitem[{{Yusifov} \& {K{\"u}{\c c}{\"u}k}(2004)}]{radial}
{Yusifov} I., {K{\"u}{\c c}{\"u}k} I., 2004, \aap, 422, 545

\end{thebibliography}

\label{lastpage}

\end{document}